\documentstyle[12pt,a4,epsf,titlepage]{article}
\begin{document}
%
%
\title{
  Lattice Dynamics of Martensitic Transformations Examined by
  Atomistic Simulations\\
  \vspace*{3.0ex} \large
  Contribution submitted to the \\  
  IV European Symposium on Martensitic Transformations \\
  \vspace*{0.5ex}
  \LARGE ESOMAT 97 \\
  \vspace*{0.5ex} \large
  July 1-5, 1997\\
  Enschede, The Netherlands  
}
\author{
  R. Meyer\thanks{electronic address: ralf@thp.uni-duisburg.de}
  and P. Entel\thanks{electronic address: entel@thp.uni-duisburg.de}\\
  Theoretische Tieftemperaturphysik \\
  Gerhard-Mercator-Universit\"at -- Gesamthochschule Duisburg \\
  Lotharstra\ss e 1, D-47048 Duisburg, Germany\\
  Tel.: ++49 - 203 - 379 - 3323 \\
  Fax.: ++49 - 203 - 379 - 2965
}
\date{
 June 23, 1997
}
\maketitle

%
%
\noindent R. Meyer \vspace*{5.6cm}
\noindent

%
%
\noindent \begin{minipage}{18cm}
  \begin{minipage}{1.5cm}  \end{minipage} \hfill
  \begin{minipage}{16.5cm} {\bf Abstract:} We have performed molecular
    dynamics simulations of ${\rm Fe_{80}Ni_{20}}$ alloys using an
    inter-atomic potential of the EAM-type which allows the simulation
    of the martensite-austenite transition. We present results,
    showing the development of an inhomogeneous shear system on a
    nanoscale during the thermally induced austenitic
    transformation. In addition to this we obtained the phonon
    dispersion relations of the martensite phase by calculating the
    dynamical structure factor from our simulation results. On
    approaching the transition temperature the phonon dispersion shows
    anomalies which might be connected with the formation of the
    microstructure during the austenitic transition. 
  \end{minipage}
\end{minipage}
\vspace*{1.1cm}

\section*{\normalsize \bf 1. INTRODUCTION}
Relationships between atomistic processes and the formation of the
microstructure during martensitic transformations are not well understood
today. In order to get more insight into this we have performed
molecular dynamics simulations of $\rm Fe_{x}Ni_{1-x}$ alloys. Within the
range $0.66 < x < 1.0$, these alloys show experimentally a
martensitic transformation from a fcc high-temperature phase to a
low-temperature bcc phase (see for example \cite{Acet} and references
therein). 

Most investigations in the context of martensitic transformations
concentrate on the 
austenite to martensite transition. This is reasonable since the
formation of microstructure during the martensitic transformation
influences the properties of the
low-temperature phase to a high degree and makes it rather difficult to
investigate the features of the homogeneous martensitic
phase. Nevertheless it is desirable to study the homogeneous
phase, in order to be able to distinguish between genuine features of the
martensitic phase and those caused by the microstructure. Therefore, we
have done simulations in the low-temperature bcc phase, looking for
the processes leading to the austenitic transition which also lead
to the formation of a microstructure in the high-temperature phase.

We present results of two distinct simulation sequences.
We begin with a study of the formation of a microstructure
on a nanometer lengthscale. These simulations required the
consideration of a large number of atoms. In order to do this we
employed a semi-empirical potential which is based on the
embedded-atom method (EAM) introduced by Daw and Baskes \cite{Daw1, Daw2}.
This enabled us to simulate systems with characteristic linear
dimension larger than 5 nm. Despite of the fact that this is,
physically spoken, still a small system, we find the formation of a
microstructure with an inhomogeneous shear system.

In a second investigation we calculated the phonon dispersion curves of
$\rm Fe_{80}Ni_{20}$ in the bcc phase. Our results show that the
austenitic transition of $\rm Fe_{80}Ni_{20}$ is accompanied by
similar phonon anomalies as those observed in many martensitic
transformations  \cite{Delaey}. We find two anomalies which we believe
to be related to the structural transition and the specific form of
the microstructure. Though pure iron also exhibits the martensitic
transformation, we demonstrate how the addition of Ni destabilizes the
bcc structure.  

\section*{\normalsize \bf 2. COMPUTATIONAL METHODS}
Classical molecular dynamics simulations have been performed, using a
semi-empirical potential based on the EAM which has been constructed
recently for the study of martensitic transformations in $\rm
Fe_{x}Ni_{1-x}$ alloys \cite{Athen, Barcelona}. In contrast to simple
pair-potentials, potentials based on the EAM are able to describe the
elastic behaviour of a metal correctly, while remaining
computationally efficient. 

Standard techniques of molecular dynamics simulations \cite{Allen}
like periodic boundary conditions and the Verlet algorithm for the
integrations of the equations of motion with a time step $ \delta t =
1.5 \times 10^{-15} s$ have been employed. Both simulation sequences
started with configurations of atoms on an ideal bcc lattice with a
random distribution of Fe and Ni atoms.

The first simulation sequence was done with a configuration of 16000
atoms ($20\times 20\times 20$ cubic bcc elementary cells) having an iron
concentration $x = 80.01$ \%. This system was heated from a temperature
of $T=600$ K to 700 K in steps of 50 K. Afterwards the temperature
was increased in steps of 20 K until the austenitic transformation
occurred at $T=860$ K. At each temperature 1000 simulation steps were
done in order to reach thermal equilibrium and another 10000 steps (15
ps) for measurement purposes. These simulations were
carried out within the isothermal-isobaric ensemble generated by the
Nos\'e-Hoover thermostat \cite{Nose, Hoover} and the Parrinello-Rahman
scheme \cite{Parrinello, Rahman} with a fluctuating simulation
box. 

The second set of simulations used a smaller configuration of $12
\times 12 \times 12$ cubic elementary cells (3456 atoms) with the same
iron concentration $x = 80.01 \%$. This system was simulated at
temperatures T = 300, 500 and 700 K with a fixed simulation box and
within the microcanonical ensemble. From a previous investigation
the equilibrium lattice parameters were known. After the equilibration
phase 40000 simulation steps (60 ps) were done, writing the atomic
positions to a file after every 10th simulation step. From this file
the dynamic structure factor 
\cite{Mermin} 
\begin{equation}
  S ( {\rm \bf q}, \omega ) = \frac{1}{N} \sum_{\rm \bf R, R'}
  e^{-i {\rm \bf q} \cdot{ ( {\rm \bf R - R' } )} }
  \int\frac{dt}{2 \pi} e^{i \omega t} \langle \exp [i {\rm \bf q
  \cdot u(R'}, 0)]  \exp [-i {\rm \bf q \cdot u(R}, t)] \rangle
\end{equation}
has been calculated ($N$ is the number of atoms and ${\rm \bf u(R,}t)$
represents the displacement of the atom from the ideal lattice position
${\rm \bf R}$ at time $t$). The phonon dispersion curves were obtained
by evaluating the positions of the peaks of $S({\rm \bf q},\omega )$
at different wave-vectors $\rm \bf q$.

\section*{\normalsize \bf 3. RESULTS}
\subsection*{\normalsize \bf 3.1 Microstructure formation}
The simulations done with 16000 atoms revealed the same general
characteristics of the martensitic phase and the austenitic transition
as similar calculations of smaller systems did\cite{Athen, Barcelona}. In
particular identical structural and orientational relationships were
observed. One of the $(110)_{\rm bcc}$ plane sets changed to a set of
close packed planes, remaining the $[001]$ direction parallel to this
planes unrotated. However, more interesting is the structure of the
resulting austenite phase. Figure 1 shows two atomic layers of the
resulting structure viewed along the unrotated $[001]$ direction. The
stacking sequence of the close packed planes turns out 
to be {\it ABCABCACABABCABCBCAC} (from left to right). Keeping in mind
the periodic boundary conditions, it can be seen from the stacking
sequence and Fig.~1 that the whole system consists of two big fcc
plates (6 and 8 layers thick) which are separated by thin 3 layer
plates. The slip faults which separate these basic blocks lead to an
inhomogeneous shear angle of $2.5^{\rm o}$. 
The atomic movements that generated the structure of the system in
Fig.~1 are elucidated by Fig.~2, which displays the deviations of the
atoms drawn in Fig.~1 from ideal body-centered lattice
positions. Comparing Fig.~1 and Fig.~2 one can see that the areas with
a regular stacking sequence are formed by a homogenous shear of the
$(110)_{\rm bcc}$ planes along the $[1\bar{1}0]_{\rm bcc}$
direction. At the slip boundaries between the homogenous regions the
displacement field reverts its direction which leads to abrupt jumps. 

\begin{figure}[p]
    \begin{center}
      \begin{minipage}{12cm}
        \epsfysize=10.5cm
        \epsffile{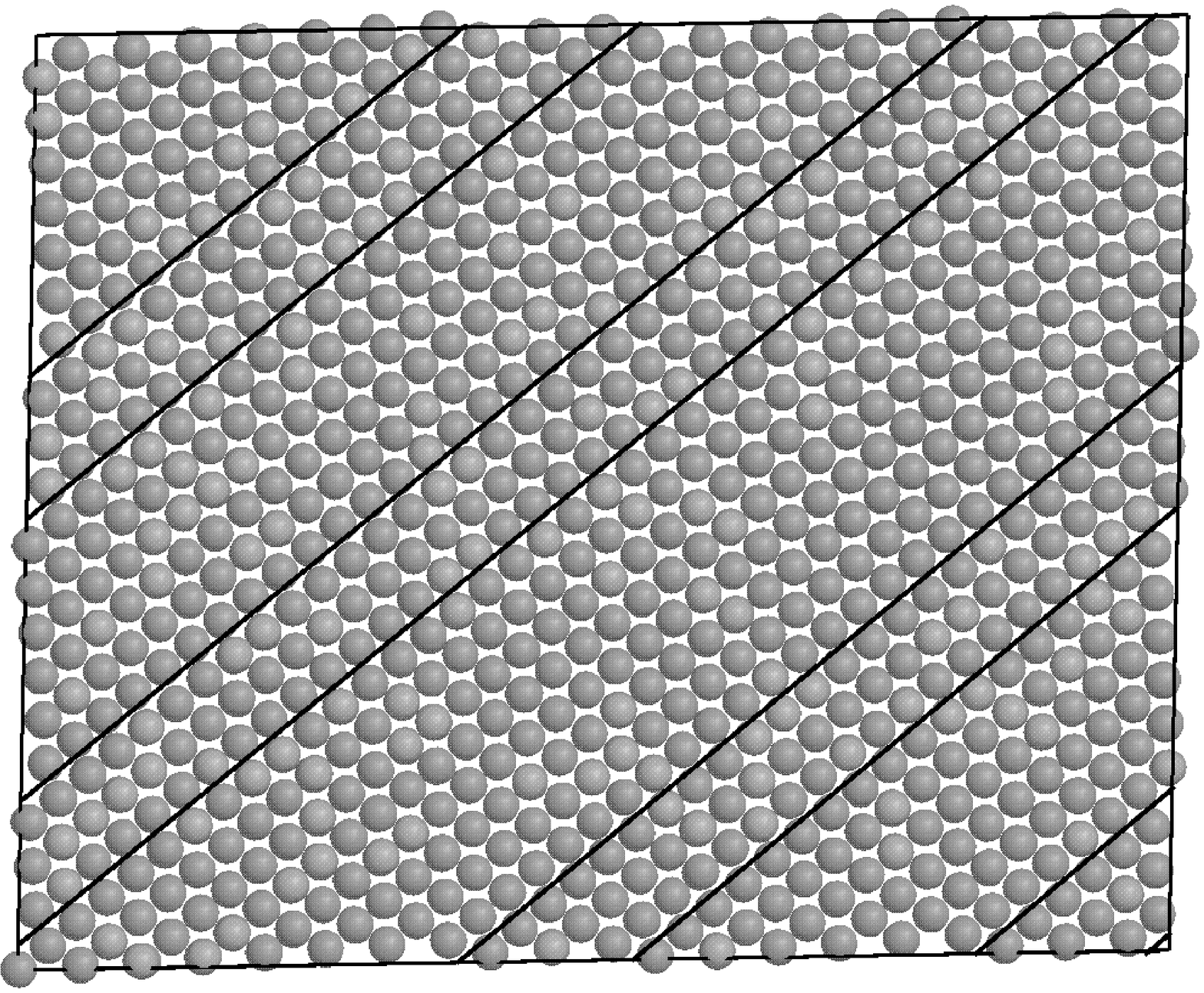}
      \end{minipage}
  \end{center}
  \xpt {\bf Figure 1:}
  Two atomic $[001]$ layers of the simulated $\rm Fe_{80}
  Ni_{20}$ crystal after the austenitic transformation. The diagonal
  lines indicate stacking faults of the $(111)$ planes.
\end{figure}

\begin{figure}[p]
  \begin{center}
    \begin{minipage}{12cm}      
      \epsfysize=10.5cm
      \epsffile{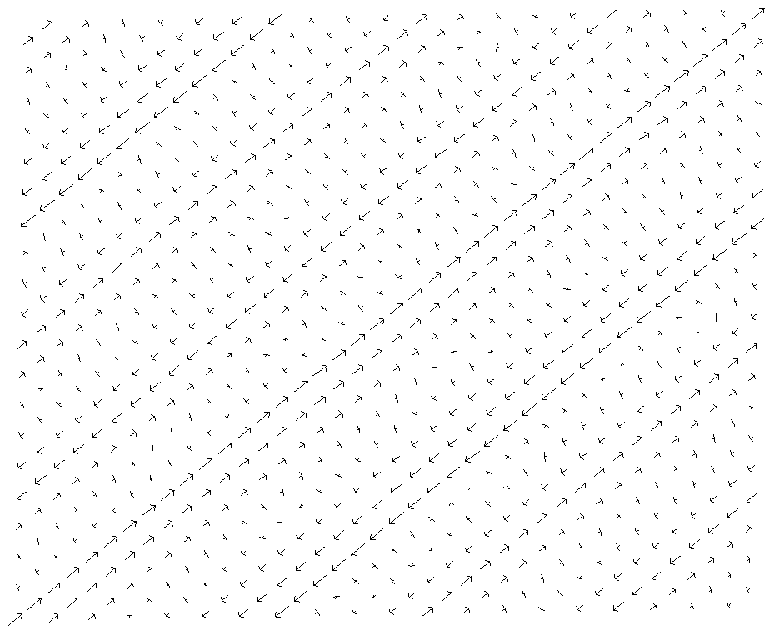}
    \end{minipage}
  \end{center}
  \xpt {\bf Figure 2:}
  Deviations of the positions of the atoms in Fig.~1
  from an ideal body-centered structure. Arrows are enlarged by a factor
  of 2.
\end{figure}

\subsection*{\normalsize \bf 3.2 Phonon dispersion relations}
In order to see the effects of alloying on the phonon
dispersion curves, first the phonon spectra of pure Fe and Ni were
calculated analytically from the EAM-potentials by diagonalization of
the dynamical matrix. Figure 3 shows the result, obtained for a
bcc structure with lattice constant $a_0 = 5.4369\,a_B$ 
determined for $\rm Fe_{80}Ni_{20}$ at $T = 300$ K from the
simulations. Along the ${\rm \bf q} = [111]$  and $[001]$ directions
both elements show a normal behaviour. Along the other directions
considered in Fig.~3 this is only true for Fe. Around 
${\rm \bf q} = [\frac{1}{2}\frac{1}{2}1]$ and along the ${\rm \bf q} = [110]$
direction the spectrum of Ni exhibits a strange behaviour and one of
the modes has negative squares of frequencies here. This result
itself is not very exceptional since experimentally Ni shows no stable
bcc phase. But the unstable mode along the
$[110]$ direction corresponds in the limit ${\rm \bf q} \rightarrow \bf 0$
to a negative elastic constant $C' = \frac{1}{2} ( C_{11} - C_{12}
)$. This is consistent with {\it ab initio} electronic structure
calculations which also reveal a negative value of $C'$ \cite{Egbert}.
\begin{figure}[t]
  \epsfxsize=18cm
  \epsffile{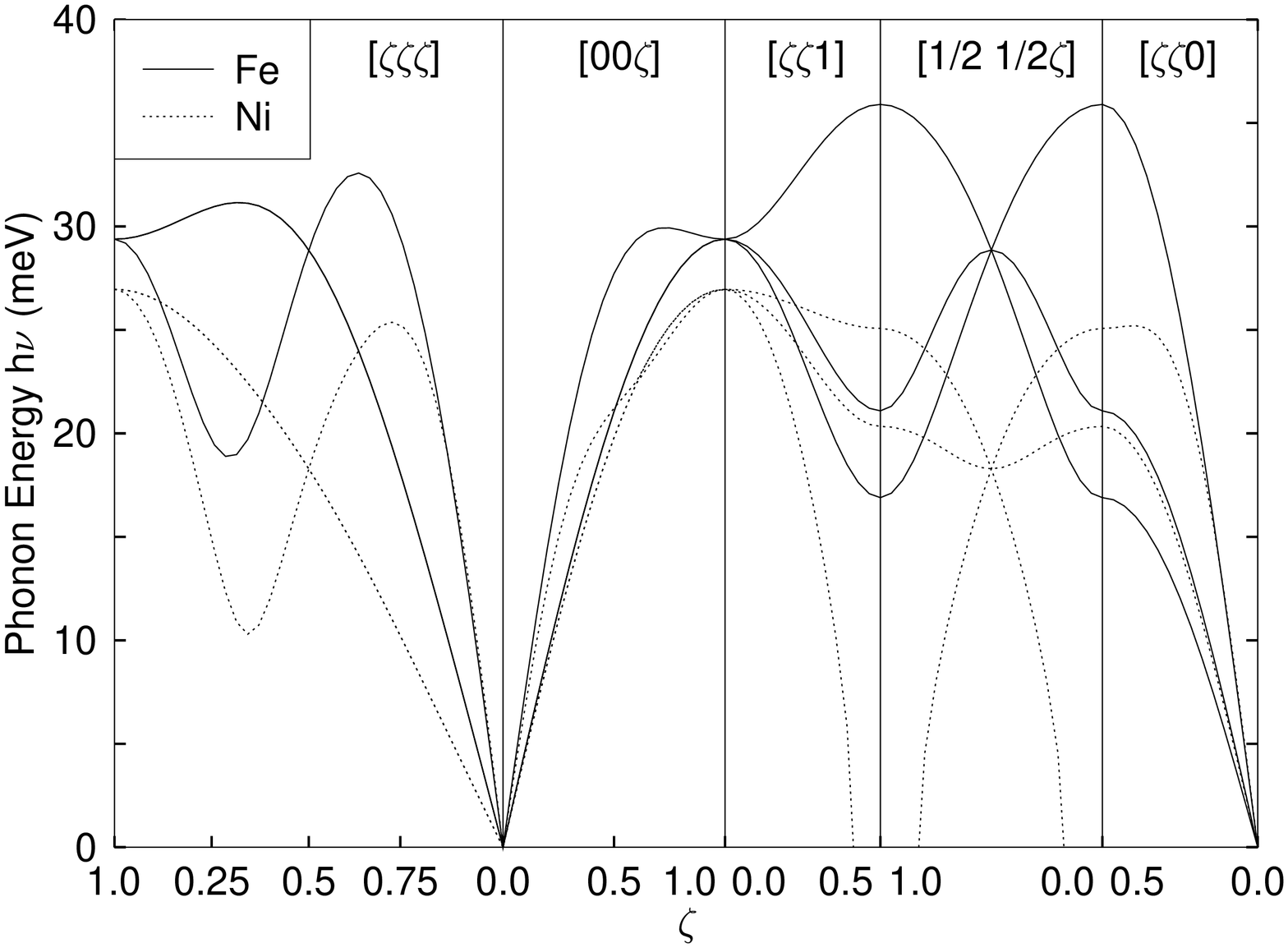}  
  \xpt {\bf Figure 3:}
  Phonon dispersion curves of bcc Fe and Ni with lattice
  constant $a_0 = 5.4369\,a_B$. The spectra have been calculated
  analytically from the EAM-potentials.
\end{figure}

With respect to this instability of Ni and the fact that the bcc - fcc
transition 
requires a shear along $[110]_{\rm bcc}$, this direction appears
to be a good candidate for phonon anomalies in those $\rm
Fe_{x}Ni_{1-x}$ alloys, which exhibit a martensitic transformation.
Figure 4 shows the phonon dispersion curves of $\rm Fe_{80}Ni_{20}$ along
this direction determined from molecular dynamics simulations by
calculation of the dynamic structure factor $S({\rm \bf q}, \omega)$
at $T = 300$ K (a), and the temperature dependence of the low lying
$\rm TA_2$ mode (b). The data points in these figures represent the 
positions of the peaks in $S({\rm \bf q}, \omega)$ averaged over all
crystallographically equivalent $[110]$ directions. The error bars are
derived from the corresponding standard deviations. If no error bars
are drawn, the standard deviations are smaller than the symbol sizes.
\begin{figure}[t]
  \epsfxsize=18cm
  \epsffile{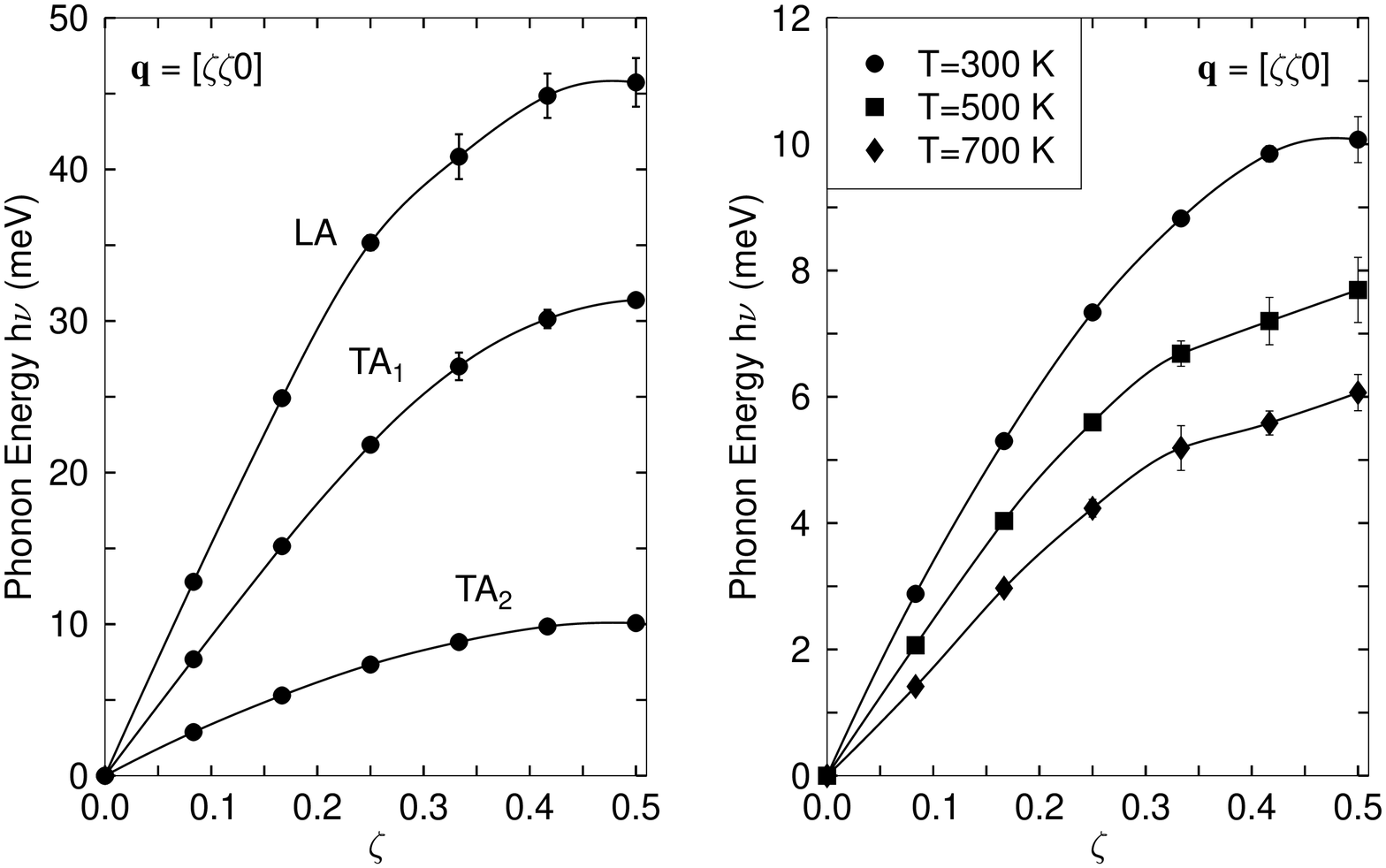}
  \parbox{8.9cm}{\begin{center} a) \end{center}} \hfill
  \parbox{8.9cm}{\begin{center} b) \end{center}}
  \xpt {\bf Figure 4:}
  (a) Phonon dispersion curves of $\rm Fe_{80}Ni_{20}$ in the bcc
  phase along $[110]$, as determined from the simulation at 300 K. (b)
  Temperature dependence of the $\rm TA_2$ mode (polarization along
  $[1\bar{1}0]$) along the $[110]$ direction.
\end{figure} 

Figure 4(a) demonstrates the dramatic effect of Ni on the phonon
dispersion curves. The longitudinal and the upper transversal modes
are shifted to higher frequencies, while the energies of the low lying
transversal branch are reduced by almost a factor 2.
At ${\rm \bf q} = \frac{1}{3}[110]$ a small dip in the longitudinal
branch occurs. But the values at this wave vector also have the largest
errors. Therefore, it is difficult to decide wether this dip together
with the errors represent physics or not.
In contrast to this in Fig.~4(b) it can be seen without doubts that
the low lying transversal mode develops two distinct anomalies on
approaching the transition temperature. At ${\rm \bf q} \rightarrow
\bf 0$
the slope of the branch reduces with increasing temperature, leading
to a positive curvature of the corresponding dispersion relation. 
The second anomaly is located between ${\rm \bf q} = \frac{1}{3}[110]$
and the Brillouin-zone boundary. The dip occurring here in the
transversal mode is also accompanied by large errors of the data points.

\section*{\normalsize \bf 4. DISCUSSION AND CONCLUSIONS}
Results of molecular dynamics simulations shown in Fig.~1 and
Fig.~2 demonstrate the formation of an inhomogeneous shear
system during the austenitic bcc - fcc transition in
$\rm Fe_{x}Ni_{1-x}$ alloys. This shear system is quite similar to
that observed in the austenite martensite - transitions of these
systems. The microstructure consists of
homogeneous fcc plates separated by slip faults. Though the length
scale of these simulations is still far too small to give definite
answers, it is interesting to see that the thicker fcc plates are
separated by rather thin plates consisting of only three atomic
layers. This could be just a random result of this particular
simulation. But this observation is confirmed by the anomalies of the
phonon dispersion curves resulting from the second simulation sequence,
since wavelength and direction of the anomaly between ${\bf q} =
\frac{1}{3}[110]$ and the zone boundary are in accordance with the
microstructure observed in Fig.~1. Therefore we think that the
formation of the microstructure during the transformation can be
attributed to this anomaly which is visible at temperatures far below
the actual transformation.
Nevertheless, simulations of even larger systems have to be done
to assure that the structures we find are not artefacts of finite-size
effects.

The other anomaly at ${\rm \bf q} \rightarrow \bf 0$, in
the phonon dispersion curves leads to a low value of the elastic
constant $C'$, which probably vanishes near the transition 
temperature. This is the driving force behind the austenitic
transformation itself and has probably no further effect on the
microstructure. From Fig.~4(a) the effects of the addition of
Nickel on the phonon dispersion curves can be seen. The most striking
effect is the reduction of the low-temperature value of $C'$. This
explains the decrease of the austenitic transition temperatures with
increasing Ni contents \cite{Acet}. 

It is clear that under normal experimental conditions, the
properties of the martensitic phase are
dominated by the microstructure which develops during the martensitic
transformation. But we think that the results of our present molecular
dynamics simulations show that the austenitic
transition also has its own genuine features which are worth to be
studied. To our knowledge there are no experimental data of the phonon
dispersion relations of the bcc phase of $\rm Fe_xNi_{1-x}$ available
at the time being. It will be very interesting to see if our
theoretical prediction can be confirmed experimentally.

%
%
\section*{\normalsize \bf Acknowledgments}
This work has been supported by the {\it Deutsche Forschungsgemeinschaft}
(DFG) within the {\it Sonderforschungsbereich} SFB 166. We also want to
thank the {\it H\"ochstleistungsrechenzentrum} HLRZ J\"ulich, Germany,
for the CPU-time on its Intel Paragon. Parts of our
calculations have been done there.

%
%

\end{document}